\documentclass [12pt,preprint]{aastex}
\usepackage{epsfig}
\usepackage{natbib}

\begin{document}
\bibliographystyle{plainnat.bst}

\voffset-0.5cm
\newcommand{\gsim}{\hbox{\rlap{$^>$}$_\sim$}}
\newcommand{\lsim}{\hbox{\rlap{$^<$}$_\sim$}}

\title{Misaligned And Alien Planets From Explosive Death Of Stars}

\author{Shlomo Dado\altaffilmark{1}, Arnon Dar\altaffilmark{1}
and Erez N. Ribak\altaffilmark{1}}

\altaffiltext{1}{Physics Department, Technion, Haifa 32000, Israel}

\begin{abstract}
Exoplanets whose orbit is misaligned with the spin of their host star 
could have originated from high-speed gas blobs, which are observed in 
multitudes in nearby supernova remnants and planetary nebulae.
These blobs grow in mass and slow down in the interstellar 
medium (ISM) by mass accretion and cool by radiation.  If their mass 
exceeds the Jeans mass, they collapse into hot giant gas planets. Most 
of the 'missing baryons' in galaxies could have been swept into such 
free-floating objects, which could perturb stellar planetary systems, 
kick bound planets into misaligned orbits or be captured themselves 
into misaligned orbits. The uncollapsed blobs can then collapse or be 
tidally disrupted into a tilted gas disk where formation of misaligned 
planets can take place. Giant Jupiters free floating in the Galactic 
ISM may be detected by their microlensing effects or by deep photometry
if they are hot.
\end{abstract} 
\maketitle

\section{Introduction}

The prevailing theories of planet formation posit that planets are 
formed in the disk of gas and dust encircling a young 
star \citep{Kuiper1951, Montmerle2006}. A 
close alignment between the rotation axis (spin) of the star and that 
of the orbital motion of the planets is expected because a star and 
its planets inherit their angular momentum from a common source - the 
protostellar disk. However, measurements of the relative spin orbit 
alignment of transiting extrasolar planets (exoplanets) through the 
Rossiter-McLaughlin effect \citep{Rossiter1924, McLaughlin1924} reveal 
that a considerable fraction 
of the hot Jupiters (exoplanets with a mass similar or greater 
than that of Jupiter and much hotter) have misaligned spin-orbit 
\citep{Hebrard2008, Winn2009a, Johnson2009, Winn2009b, Winn2009c, 
Pont2009, Pont2010, Narita2009, Narita2010, Winn2010, Schlaufman2010, 
Hirano2010, Triaud2010, Hebrard2011, Simpson2011} . 
Other properties of hot Jupiters, such as a temperature 
independent of 
their distance from their host star and of its temperature as observed 
for the hot Jupiters orbiting the nearby star HR 8799 
\citep{Marois2010}, also challenge \citep{Close2010} the {\it in 
situ} planet formation theories \citep{Mizuno1980, Nero2009}. 

Native planets formed in the protostelar disk may be scattered later into 
misaligned orbits by planetary encounters \citep{Chatterjee2008, Ford2008} 
or by the Kozai mechanism \citep{Malberg2007, Fabrycky2007}. But, such 
mechanisms are unlikely to explain the large fraction of hot Jupiters with 
misaligned orbits \citep{Hebrard2011} and a temperature not correlated with 
their distance from the host star \citep{Marois2010}. Here we suggest an 
alternative plausible {\it ex situ} origin of hot giant gas planets with 
misaligned spin-orbit, namely, the formation of high-speed gas blobs in the 
explosive death of stars. Such gas blobs are observed in large numbers in 
nearby supernova remnants \citep{Fesen2006, Fesen2007} planetary nebulae 
\citep{O'Dell2002, Matsuura2009} and star formation regions. The existence 
of these gas blobs with cometary-like appearance (see Figs.1,2), probably 
due to overtaking winds, was not expected, their origin is still not 
understood and their fate is unknown. We shall argue that they grow in mass 
and slow down in the interstellar medium (ISM) by sweeping in the ambient 
matter in their way. They cool by radiation and if their mass exceeds the 
Jeans mass \citep{Jeans1902} they contract gravitationally into giant  
Jupiters. The collapsed and uncollapsed Jupiter-mass objects can perturb 
stellar planetary systems and kick bound planets into misaligned orbits. 
They can also be captured into misaligned orbits around host stars. The 
uncollapsed ones may then collapse or be tidally disrupted and form a 
tilted gas disk where {\it in situ} formation of planets with misaligned 
spin-orbit can take place. 

If the numbers and properties of the cometary blobs (CBs) observed in the 
supernova remnant Cassiopeia A (Cas A) and in the Helix nebula represent 
faithfully those formed in other supernova remnants (SNRs) and planetary 
nebulae (PNs), then the total number of Jupiter-mass objects free-floating 
in the ISM of our Galaxy may exceed the number of stars by more than two 
orders of magnitude. These free-floating Jupiter-mass objects could 
have accreted most of the gas and dust in the ISM of galaxies and may 
explain the presently small mass ratio between gas and stars in galaxies 
($\sim$ 1:10 in the Milky Way).

The fraction of CBs that end up as free-floating Jupiters in the 
Galaxy cannot be reliably estimated from theory. However, the density of 
free-floating Jupiters in the Galaxy can be measured through their 
microlensing effects. In fact, presence of a significant fraction of the 
'missing baryons' in galaxies in free-floating Jupiters is 
consistent with the measured optical depth for microlensing towards the 
Galactic bulge by the MACHO \citep{Alcock2000a} and OGLE \citep{Sumi2006} 
collaborations (see section 5). The MACHO collaboration also reported an 
optical depth towards the stars in the Large Magellanic Cloud (LMC) which 
is larger than those expected from lensing by the known populations of 
stars in the Galaxy and the LMC \citep{Alcock2000b, Bennett2005}. However, 
the microlensing optical depths towards the LMC stars that were measured 
by the EROS \citep{Tisserand2007} and OGLE \citep{Wyrzykowski2010} 
collaborations, are much smaller than that measured by MACHO and are 
consistent, within errors, with those expected from the known stellar 
populations in the LMC and the Galaxy. In view of the conflicting results, 
the small statistics of lensing events, and the large uncertainties in the 
modeling of the LMC, it is difficult to draw a definite conclusion on 
free-floating planets in the LMC from the measured values of the 
microlensing optical depths of the LMC. Second generation microlensing 
surveys using ground based telescopes, or space based telescopes such as 
The Wide-Field Infrared Survey Telescope (WFIRST), that unlike Kepler, 
will be sensitive also to free-floating planets \citep{Bennett2011, 
Catanzarite2011} and will measure or set limits on their density in the 
ISM, their masses and the fraction of baryons in the Galaxy which reside 
in such planets.

Observational support for the hypothesis of {\it ex situ} 
origin of misaligned hot Jupiters and other planets in 
orbits around host stars could also come from the detection of 
a large population of free-floating 
hot Jupiters in the ISM in deep pencil searches with extremely large 
optical and infra-red telescopes (ELTs) equipped with advanced 
adaptive optics and having a much greater light collecting ability 
than that of the current largest telescopes. A large population of 
free-floating hot Jupiters in the ISM may also contribute 
significantly to the unresolved diffuse infra-red background 
radiation.

Finally, dramatic flux changes occurring over several weeks to months in 
the radio flux from quasars, such as 0954+658 \citep{Fiedler1987}, and 
episodes of multiple images of radio pulsars \citep{Cordes1986, 
Rickett1990} were shown \citep{Walker1998} to be reproduceable by 
refraction from the ionized skin of an hypothesized large population of 
neutral clouds having a few astronomical units in radius and masses $\lsim 
10^{-3}\, M_\odot$, which are free floating in the Galactic ISM and are 
exposed 
to the Galactic ionizing UV flux.

\section{Cometary Blobs in SNRs and PNs} 

High resolution imaging of nearby young supernova remnants from the 
explosive death of massive stars, such as SNR Cass A \citep{Fesen2006} and 
SNR 3C 58 \citep{Fesen2007} and of nearby planetary nebula remnants of the 
explosive death of sun-like stars, such as the Helix nebula 
\citep{Matsuura2007, Matsuura2009}, the Ring nebula, the Dumbbell nebula 
and the Eskimo nebula, reveal that thousands of cometary blobs (CBs) with 
masses similar to that of Earth and radii of hundreds of astronomical unit 
are moving away from the center of the explosion along radial directions.  
The fate of CBs formed in the explosive deaths of stars is not known. 
Probably, most of these CBs grow in mass and slow down by accreting the 
ISM particles (gas and dust) in their path, and cool by radiation. When 
their mass exceeds the Jeans mass \citep{Jeans1902}, $M_J\!=\!3\,k\, 
T\,R/4\, G\,m_p$, where $k$ is the Boltzman constant, $T$ is the 
temperature of the $H_2$, $R$ is the radius of the CB and $m_p$ is the 
proton mass, they collapse gravitationally into a hot Jupiter.

The outer 1825 CBs observed in the SNR Cass A \citep{Fesen2006} have 
typical velocities $V_0\!\approx\!10,000$ km/s, a radius $R\!\approx\! 
0.1''$ corresponding to 0.002 pc (at 3.4 kpc, the estimated distance 
to Cas A), and $n_b\!\geq\! n_e\!=\! 0.7\times 10^4\, {\rm cm^{-3}}$, 
i.e., a mean column density $N_{CB}\! \sim\!6\times 10^{19}\, {\rm 
cm^{-2}}$, and a mass $\!\geq\!1.2\times 10^{28}$ g ($\!\sim\!$ twice 
that of Earth).

The measured radial velocities of over 450 CBs in SNR 3C 58, reveal 
\citep{Fesen2007} 
two distinct populations, one with average projected velocities of 770 
km/s, and a high NII/H$_\alpha$ line emission ratio and the other showing 
velocities less than 250 km/s and a much lower NII/H$_\alpha$ line emission 
ratio. It was suggested that the low velocity populations were formed in 
the ejecta of the progenitor star long before its supernova explosion while 
the higher velocity CBs were formed during or after the SN explosion.

The estimated current rate of Galactic core collapse SN explosions 
\citep{Tammann1994} is $\sim\!$ 1/50 y. This rate, being proportional to 
the 
Galactic star formation rate, was much higher in the past. Thus, over the 
past 13.2 Gy or so, the estimated age of the oldest stars in the Galaxy
\citep{Ferbel2007}, there have been  more than $2.6\times 10^8$ core 
collapse SN 
explosions. If the mean number of CBs formed in an SN explosion is 
represented by the 1825 CBs resolved in the SNR Cas A, then more than 
$10^{12}$ CBs were launched into the ISM of the Galaxy in SN explosions 
during its age.

Many more CBs are launched in planetary nebulae than in SN explosions. 
Detailed observations of the nearby Helix nebula at a distance of 219 
pc with high resolutions telescopes such as the Hubble Space 
Telescope (HST), the Very Large Telescope (VLT) and the Subaru 
telescope revealed more than 40,000 CBs \citep{Matsuura2009}. If these 
CBs are well 
represented by the CB KI, which was studied in detail in 
Matsuura et al.~2007, then 
their typical radius is 0.75 arcsec or $R$=0.0008 pc and their H$_2$ 
density is $8\times 10^4\, {\rm cm^{-3}}$. These yield a CB mass 
$1.7\times 10^{28}$ g, which is three times larger than the mass of 
Earth, ($5.98\times 10^{27}$ g) and a mean column density $N\!\sim\! 
2\times 10^{20}\, {\rm cm^{-2}}$. 

The local space density of white dwarf stars, 
presumably mostly born in PNs, was found to be \citep{Holberg2002} 
$\,\sim\! 
5.0\times 10^{-3}\,{\rm pc^{-3}}$, with a corresponding mass density 
of $\!\sim\! 3.4\times 10^{-3}\,M_\odot\, {\rm pc^{-3}}$, which is 
roughly 5\% of the local mass density in stars that was 
measured \citep{Creze1998}
with Hipparcos ($\!\sim\! 7.6\times 10^{-2}\,M_\odot 
\, {\rm pc^{-3}}$ corresponding to a local number density 
$n_*\!\sim\! 0.23 \, {\rm pc^{-3}}$). The birth rate of planetary 
nebulae is roughly that of white dwarfs. Assuming that the number of 
CBs per planetary nebula is $\sim\! 40,000$ as was observed in the 
Helix Nebula with the Spitzer space telescope and the Subaru telescope 
(see Fig.1) and that the local ratio of white dwarfs to stars 
represents the Galactic ratio, then the local density of Jupiter-mass 
objects is roughly $n_{_J}\!\sim\! 200\, {\rm pc^{-3}}$ and their 
number in the Galaxy exceeds the number of stars by roughly three 
orders of magnitude.

\section{The fate of high speed CBs}

High-speed projectiles such as those ejected in SN explosions and 
launced by microquasars decelerate in the ISM by accreting the gas and 
dust on their way. 
Momentum conservation with the neglect of radiative losses seems to
describe well their deceleration: Consider
first the deceleration of the spherical shell of initial mass, radius
and radial velocity, $M_0\,, R_0\,,V_0$, respectively, ejected 
in SN explosion into an ISM of a constant density $n_{ism}$.
Momentum conservation with the neglect of energy losses 
yields a radial velocity 
$V\!=V_0/(1\!+\!\Delta M/M_0)$, where
as a function of swept in ISM mass 
$\Delta M\! =\! 4\,\pi\, n_{ism}\,m_p\,
(R^3\!-\!R_0^3)/3$.  
In the limit $R\!\gg\! R_0$, the age as a function of 
the radius of the expanding shell has the form
\begin{equation}
t={R\over V_0}\, \left[1+
{\pi\ R^3\, n_{ism}\, m_p\, \over 3\,M_0}\right]\,.
\label{tSN}
\end{equation}
Eq.~(\ref{tSN}) describes well the transition from a
linear expansion of the entire SN shell, $R\!\approx\! 
R_0\!+\!V_0\,t$, 
observed, e.g. in 
SN1987A  and SN1993J at early times to the asymptotic behaviour,
$R\! \approx\! [3\, M_0\,V_0\, t/\pi\,n_{ism}\,m_p]^{1/4}$
for swept in mass $\Delta M\! \approx\!4\, \pi\,n_{ism}\,m_p\, R^3/3 
\!\gg M_0/4$.

The slow-down of a CB of initial mass $M_0$, radius $R$, baryon 
density $n_b$ and velocity $V_0$ by mass acreetion along its path in the 
ISM is described by $M\,dV\!=\!-\!V\,dM$, i.e., 
\begin{equation}
{M_0\, V_0\over V}\,dV =-V^2\, \pi\, R^2\, n_{ism}\, m_p\, dt.
\label{deceleration}
\end{equation}
In order to reduce its velocity by a factor $k$, 
a CB must accrete an ISM mass $\Delta M\!=\! (k\!-\!1)\, M_0$.
Neglecting its relatively slow
thermal expansion, it 
must sweep in a column density $k$-1 
times its own mean column density $N_{CB}\!=\!4\, n_b\, R/ 3$.
i.e., cross a
distance $d\!=\! (k\!-\!1)\,(4\, n_b\, R/ 3\, n_{ism})$
which takes a time $t_d\!=\!(2\, n_b\,R/3\, n_{ism}\,V_0)\, (k^2-1)$.

Initially, the temperature of the surface of a CB facing the young neutron 
star (ns) in SNRs or the white dwarf (WD) in PNs is determined by its 
illumination by the young neutron star at the center of the SNR or the 
white dwarf at the center of the PN. For instance, in the absence of an 
internal or an external heat source other than the light of the white 
dwarf at the center of the PN at a distance $D\!\sim\! 0.8$ pc, their 
surface temperature $T_{CB}$ of their illuminated side would be roughly, 
$T_{CB}\!<\! T_{WD}\, \sqrt{R_{WD}/2^{1/2}\,D}\!\sim\!0.42$K for a typical 
WD radius $R_{WD}\!\sim\! 7000$ km and WD surface temperature 
$T_{WD}\!\sim\! 30,000$K. The illuminated surface, however, is visible 
from Earth at such distances by light emitted from the decay of atomic and 
molecular levels which are excited by the incident light. A thin 
photoionized "skin" of the CBs can be maintained by the Galactic UV 
background radiation even at large distances from their source. This 
photoionized skin of CBs in the ISM can be responsible for strong 
scintillations observed in the radio emission from extragalactic and 
Galactic compact radio sources \citep{Walker1998}.

Far away from the central star, the only significant  heat source 
of CBs is the kinetic energy 
deposited by the collision of the swept in ISM gas and dust 
particles with the CBs' molecular gas. 
Without any other heat source, the CBs cool 
by radiative decay of rotational and vibrational molecular levels
excited by internal collisions.  
The temperature of a CB adjusts itself such that its
cooling rate by radiation equals the rate of this energy deposition.
If the radiation can be approximated by a black body radiation, then
\begin{equation} 
4\, \pi\,R^2\, \sigma\, T^4={n_{ism}\, m_p\,V^3\over 2}\, 
\pi\,R^2\,.  
\label{equilibrium}
\end{equation}
This yields a rather low equilibrium temperature 
\begin{equation} 
 T=\left[{n_{ism}\, m_p\,V^3\over 8\,\sigma}\right]^{1/4}
 \approx 0.044\, \left[{V\over {\rm km}}\right]^{3/4}\, 
 \left[{n_{ism}\over {\rm cm^3}}\right]^{1/4}\, {\rm K}\,,
 \label{T}
\end{equation}
where $\sigma$ is the Stefan-Boltzman constant. This equilibrium
temperature decreases when the CB decelerates by mass accretion
from  $\sim$ 44K for  $V\!=\!10,000\, {\rm km\,s^{-1}}$,
to below 10K for $V\!<\!1400\, {\rm km\,s^{-1}}$.

Because of their low temperature and their large size, thermal 
expansion of CBs is negligible during their deceleration: 
Consider a non collapsed spherical blob of $H_2$ gas of a total mass
$M$, a constant density and an initial radius
$R_0\!=\!10^{15}$ cm that expands with the speed of sound $\dot
R\!=\!c_s\!=\!\sqrt{\gamma\, k\, T/2\, m_p}$, where $\gamma$=1.4 is the
adiabatic constant for an ideal H$_2$ gas. If the CB cools mainly by  
radiation according to the Stefan Boltzman law, then 
$(3/2)\,(M/2\,m_p)\, k\,
\dot{T}\!=\!-\!4\,\pi\,R^2\, \sigma\,T^4$. By dividing these two rates,
separating the $R$ and $T$ dependences and integrating the resulting
equation from initial $T_0$ to a final $T\!\ll\!T_0$, we obtain
\begin{equation}
{R-R_0\over R_0} \approx R_0^{-3}\,  \sqrt{{\gamma\, k\,T\over 2\,
m_p}}\, {3\over 40}\,{M\over m_p}\, {k\, T\over 4\, \pi \sigma\,
T^4}\, ,
\label{exp}
\end{equation}
where we have assumed that $(R\!-\!R_0)\!\ll\! R_0$. Indeed, for
$M\!<\!0.01\,M_\odot$ and
$T\!\sim\! 10$K, the RHS of the last equation yields
a negligible expansion, $(R\!-\!R_0)/R_0\approx 0.0035$.

The outer CBs observed in the SNR Cas A \citep{Fesen2006} have 
typically a 
mean column density $N_{CB}\! \geq\!6\times 10^{19}\, {\rm cm^{-2}}$, 
and a mass 
$\!\sim\!1.2\times 10^{28}$ g ($\!\sim\!$ twice that of Earth).  Hence 
Galactic column densities above a few $10^{21}\, {\rm cm^{-2}}$ are 
sufficient to slow down the CBs from Cas A to velocities smaller 
than the local Galactic escape velocity \citep{Smith2007}  
$V_{esc}\!\sim\!550\,{\rm 
km/s}.$ Such slowed-down CBs continue to accrete interstellar matter, grow 
up in mass, slow down further and cool by radiation until they are 
captured by a host star or their mass exceeds the Jeans mass, $\!\sim\! 
0.01\,M_\odot$ for $T\!\sim\!10$K and $R\!\sim\! 10^{15}$ cm, and they 
collapse to hot Jupiters free-floating in the Galactic ISM. 
Moreover, 
most star formation and SN explosions take place within large molecular 
clouds. This is a natural consequence of their low temperatures and high 
densities. Large molecular clouds have a typical size of tens of pc and a 
typical baryon density $n_b\!\sim\! 10^{3}\, {\rm cm^{-3}}$. Most of 
the CBs which 
are ejected in a supernova explosion within such a dense environment slow 
down completely within the molecular cloud. Some may escape into the ISM 
with rather a small velocity and virialize there.  Some of the CBs in the 
molecular clouds can seed there star formation by mass accretion, while 
other CBs can be captured there into planetary orbits around newly born 
stars. Very high velocity CBs from SN explosions outside molecular clouds 
that are moving in Galactic directions of small column density will escape 
into the low-density intergalactic space where they will expand before they 
reach the Jeans mass and enrich the intergalactic medium (IGM)  with 
metals produced in SN explosions.

In the case of SNR 3C 58, the measured radial velocities of the CBs 
are smaller than those observed in Cas A, roughly by an order 
of magnitude, probably because of their deceleration during the long time 
since the SN explosion (about 3000-4000 yeas ago) relative to that of 
Cas A (about 330 years ago).

The fate of the CBs observed in planetary nebulae probably is similar. 
If they are well represented by CB K1 in the Helix nebula at a 
distance of 219 pc that  has been studied in detail  
\citep{Matsuura2007, Matsuura2009}, they  have 
a typical CB mass of $1.7\times 10^{28}$ g, which is three times 
larger than the mass of Earth, and a mean column density $N\!\sim\! 
2\times 10^{20}$. Their mass grows by collision and merger with other 
CBs and by accreting the ISM gas and dust in their way. They cool by 
radiation until they are captured by a host star or their mass exceeds 
the Jeans mass and then they collapse gravitationally. After collapse 
they accrete rather a negligible amount of additional ISM gas and dust
because of their small size.

A large fraction of the Jupiter-mass objects will end in the 
Galactic disk as uncollapsed Jupiter-mass gas clouds or free-floating 
Jupiters. Some of these Jupiters may be captured through dynamical 
friction, first into the Oort cloud of stars and from there into 
misaligned and eccentric planetary orbits. They can also kick native 
planets in aligned orbits around a star into misaligned orbits, 
particularly if they are super Jupiters. The encounter rate of an alien 
planet with the Oort cloud of stars is given roughly by $n_{_J}\, \sigma\, 
v\!\approx\! 10^{-14}\,{\rm s^{-1}}$ where we adopted a 1000 astronomical 
units as the effective radius of the Oort cloud and a virial velocity 
$v\sim 30\, {\rm km\, s^{-1}}$ of alien planets. Such a rate yields
thousands of encounters with a star during an age comparable to the age of 
the Galaxy.

Uncollapsed free-floating gas clouds can also be captured by stars and be 
tidally disrupted into a disk with a misaligned 
spin-orbit around the host star, or strongly perturb their planetary system.
Their star crossing rate is roughly $n\, \sigma\, 
v\!\approx\!10^{-16}\, {\rm s^{-1}}$ where $n\!\sim\! 200 {\rm 
pc^{-3}}$ is their local density, $\sigma \!\sim\!\pi\, 10^{30}\,{\rm 
cm^2}$ is 
their cross section and $v\sim 30\, {\rm km\, s^{-1}}$ is their 
velocity. 
Such rates yield a high probability of  
Jupiter-mass clouds to be captured or strongly perturb the 
planetary system of most of  
the stars of the Galaxy during their life.
Formation of planets in such misaligned disks around host stars
or perturbing strongly its planetary system will result in 
planets with eccentric and  misaligned spin-orbit.

\section{Observations of alien protoplanetary disks?} 

The prevailing theories of planet formation posit a protoplanetary 
disk which is part of the protostellar disk left over after the 
formation of the host star. Such theories imply a close alignment 
between the rotation axis of the star and that of the protoplanetary 
disk. However, alien disks which are formed by capture of cometary 
blobs are expected to be warped and misaligned. Employing adaptive 
optics or interferometry it is possible to image some of the 
planet-forming disks in nearby systems and determine their alignment 
\citep{Watson2010}. While in some of the systems the disk is 
symmetrical, in others \citep{Greaves1998, Buenzli2010, Kloppenborg2010}
it is clearly lopsided, which hints at processes occuring 
much later than the original stellar formation. What we might be 
seeing in these systems is the recent capture of a cometary blob, 
which may form misaligned planets.

\section {Free-floating planet detection through microlensing} 

Microlensing of stars \citep{Liebes1964} by planets has already 
been used to search  for free-floating  planets \citep{Quanz2010}.
The effective cross section for gravitational microlensing  
of a star at a distance $D_s$ by an intervening mass
$M$ at a distance $D_l$ is given roughly by 
$\sigma_l(D_l)\!=\!\pi\,\theta_E^2\, D_l^2$,
where 
\begin{equation}
\theta_{_E}=\sqrt{4\,G\,M\, (D_s-D_l)\over c^2\, D_s\,D_l}
\label{Ering}
\end{equation} 
is the angular radius of the Einstein ring image of the lensed star 
created if the lens is lying on the line of sight to the star.

The optical depth for gravitational lensing
of a star in the Galactic bulge at a distance $D_s$
by free-floating Jupiters with 
a mass density $\rho_{M}[x]$ 
at a distance $D_l\!=\! x\, D_s$  along the line of sight to the
star is given by, 
\begin{equation}
\tau={4\,G\,D_s^2\over c^2}\, \int_0^1 \rho_{M}[x]\,x\, (1-x)\, dx\,.  
\label{taul}
\end{equation}
Thus, if the mass density of planets is 
proportional to the stellar mass density, then they
contribute to the total optical depth for microlensing
in the same proportion. It follows from Eq.~(\ref{taul}) that the 
optical depth for microlensing of stars  
in the Galactic bulge by stars along their line of sight is given 
approximately by  
\begin{equation}
\tau_{bulge}\approx {R_{Sch}(M_*)\over 4\,\pi\, h_r}
\approx 1.7\times 10^{-7}\,,
\label{taubulge}
\end{equation}
where $R_{Sch}(M_*)\!=\!2\,G\,M_*/c^2\!\approx\! 3.5\, M_*/M_\odot$ km 
is the  Schwarzschild radius of the total mass of the stars in the Galaxy, 
$M_*\!\approx\! 5\times 10^{10}\, M_\odot$, and 
$h_r\!\approx\!2.7$ kpc is the stellar scale length in the Galactic 
disk. This optical depth is much smaller than $\tau_{bulge}\!=\!(3.23\!
\pm\!0.50)\times 10^{-6}$  measured by the MACHO collaboration 
\citep{Alcock2000a} and  $\tau_{bulge}\!=\!(4.48\!\pm\!2.37)\times 
10^{-6}$ measured by the OGLE collaboration \citep{Sumi2006}. These 
measurements leave 
enough room for a significant contribution from free-floating Jupiters 
to the microlensing optical depth towards the Galactic bulge.

The typical duration of a microlensing event 
is the lens crossing time of the Einstein ring,  $\Delta t\!\sim\! 
\theta_{_E}\,D_l/V_t$,
where the $V_t$ is the lens-source relative transverse velocity observed 
from Earth.
The typical duration of microlensing events of stars in the Galactic bulge 
($D_s\!\approx\!  8$ kpc) by free floating Jupiters ($M\!\sim\! M_j 
\!\approx\!0.95\times 10^{-3}\, 
M_\odot$) in the  Galactic bulge  ($D_s\!-\!D_l\!\sim\! 1$ kpc)
and for $V_t\!\sim\!120$ km/s 
(the typical random velocity of stars in the Galactic bulge) is roughly,
\begin{equation}
\Delta t \approx \sqrt{{4\,G\,M_j\, (D_s-D_l)\over c^2\, V_t^2}}\approx
1\,{\rm day}\, .
\label{tcrossing}
\end{equation} 
The same lens placed at  half the distance to the bulge 
has the maximal effective cross section for microlensing of bulge 
stars  that is only larger roughly by a 
factor 2. The duration of the microlensing event is similar, 
$\Delta t\!\sim\!1$ day. However, if 
the distribution of free 
floating Jupiters is proportional to that of the Galactic stars,
then most of the micro lensing events of bulge stars 
by floating Jupiters are by those 
present in the  bulge, with a typical duration of 1 day.  

For the LMC, where 
$M_*\!\approx\!5.3\times 10^{9}\,M_\odot$ and $h_r\!\approx\!1.6$ kpc
\citep{Alves2000},
Eq.~(\ref{taul}) yields an optical depth for self lensing of LMC 
stars by LMC stars, $\tau_{LMC}\!\approx\!1.1\times 10^{-8}$,
while the MACHO collaboration reported \citep{Alcock2000b, Bennett2005}
$\tau_{bulge}\!\approx \!(1.0\!\pm\!0.3)\times 10^{-7}$,  
the  EROS collaboratin found \citep{Tisserand2007}
$\tau_{LMC}\!<\!3.7\times 10^{-8}$, 
and the OGLE collaboration concluded  \citep{Wyrzykowski2010} 
that $\tau_{LMC}\!\approx \!(1.6\!\pm\!1.2)\times 10^{-8}$.
The spread of the measured values of the microlensing optical depth 
towards the LMC stars and the large error bars also leave enough 
room for a significant contribution from free-floating Jupiters
in the LMC to the microlensing optical depth of the LMC.

Although microlensing surveys seem to be promising for the search for 
free-floating Jupiters, they cannot detect uncollapsed Jupiter-mass clouds 
if their density distribution is such that their enclosed mass within a 
radius $r$ satisfies $M(<r)\approx (r/R)\,M$ (e.g., an isothermal sphere). 
The Einstein radius of such uncollapsed clouds (i.e., clouds with masses 
smaller than $\!\sim\! 0.001M_\odot$) placed anywhere along the line of 
sight to the Galactic bulge is smaller than their radius of $\!\sim\! 
10^{15}$ cm, at least by two orders of magnitude, which makes them 
practically nondetectable through gravitational lensing. However, evidence 
for a large Galactic population of free floating clouds with masses 
$M\!\lsim\! M_J$ might come from future 
observations of scintillations in the 
radio lightcurves of galactic and extragalactic compact radio sources such 
as pulsars, microquasars, quasars and gamma ray bursts.
Future ifrared lensing surveys might alao reveal free-floating Jupiter 
mass objects magnified by stars.

\section{Evidence from scintillations of compact radio sources}

Dramatic flux changes occurring over several weeks to months in the radio 
flux from quasars \citep{Fiedler1987} such as 0954+658 and episodes of 
multiple images of radio pulsars \citep{Cordes1986, Rickett1990} were 
shown \citep{Walker1998} to be reproducible by refraction from the 
ionized skin of an hypothesized large population of neutral clouds having 
a few astronomical units in radius and masses 
$\lsim\!M_J\!\approx\!10^{-3}$ Msun, which are free floating in the 
Galactic ISM and are exposed to the Galactic ionizing UV flux.

\section{Conclusions}

Our paper proposes a plausible common solution to two important 
astronomical puzzles, namely, the fate of the missing gas in galaxies 
and the origin of misaligned planets: Most of the baryons in galaxies 
may reside in Jupiter-mass objects free-floating in the ISM and not in 
stars. This may explain whereto most of the gas in galaxies has 
disappeared \citep{Fukugita2004}. The capture of such objects by host 
stars or their interaction with the planetary system of the host stars 
may be the main origin of misaligned planets and hot Jupiters.

Modelling formation of planets by gravitational collapse is difficult, in 
particular the capture of free floating clouds by host stars followed by 
planet formation. In fact, so far numerical simulations, disputably, have 
not been fully successful in reproducing planet formation through 
gravitational instability within or without protoplanetary 
disks\citep{Durisen2005, Boley2009, Dodson-Robinson2009, Rafikov2011}. 
Observations may be a more promising route for testing the proposed {\it ex 
situ} origin of misaligned planets.

Compelling observational support for an {\it ex situ} origin of misaligned 
planets and hot Jupiters in distant orbits may come from the detection of 
a large population of free-floating hot Jupiters in the ISM in deep pencil 
searches with extremely large optical and infra-red telescopes (ELTs) 
equipped with advanced adaptive optics and having a much greater light 
collecting ability than that of the current largest telescopes. A large 
population of free-floating hot Jupiters in the ISM may also contribute 
significantly to the unresolved diffuse infra-red background radiation.

Free-floating planetary mass objects in the Galactic ISM and the Galactic 
halo may also be discovered through gravitational microlensing. They 
may explain why both the MACHO and OGLE surveys
have found a total microlensing optical depth towards the Galactic 
bulge higher than predicted by contemporary Galactic models.  The 
translation of the MACHO and OGLE results to a density of 
free-floating planet-mass objects however is sensitive to the unknown 
mass function of low-mass stars (red and brown dwarfs) and planets.

Future microlensing surveys will be more sensitive to free-floating 
planets, in particular space based surveys with telescopes such as 
the Wide-Field Infrared Survey Telescope (WFIRST)
that unlike Kepler will be sensitive 
also to unbound planets. These new projects together with second 
generation of ground based microlensing surveys may discover 
a large population of free-floating 
planets in the Galactic ISM and determine what fraction of the missing 
baryons resides in such planets.

Evidence for a large population Jupiter mass clouds free floating in the 
Galactic ISM may come from radio and/or optical 
scintillations of Galactic and extragalactic sources
\citep{Walker1998, Draine1998}.

Alien planets, whether directly captured or formed {\it in situ} from an 
alien disk, can be distinguished by their unusual age, misaligned or 
eccentric orbit, or chemical composition. Earlier on, they can be observed 
as a warped gas and dust cloud caught soon after the action of collision 
or capture by a host star.

{\bf Acknowledgment:}
We thank Dan Maoz and Noam Soker and an anonymous referee for useful 
comments.

{\bf Note added after submission of the paper for publication in ApJ}\\ 
The detection of a large Galactic population of unbound Jupiter mass 
objects, roughly twice as many as the number of main sequence stars with 
mass $M\!1>\!0.08\!M_\odot$, through microlensing observations of stars in 
the Galactic bulge was reported in Nature on May 19, 2011 by the 
Microlensing Observations in Astrophysics (MOA) collaboration and the 
Optical Gravitational lensing Experiment (OGLE) \citep{Sumi2011}. The 
existence of 
such a large Galactic population of free-floating Jupiter-mass objects was 
suggested in this paper, which was submitted for publication in ApJ on 
February 13, 2011, as supportive evidence for the proposed ex-situ origin 
of misaligned and alien planets (fast CBs that are produced in Galactic SN 
explosions, but slow down sufficiently to avoid escape into the 
intergalactic space).

\begin{figure}[]
\centering
\epsfig{file=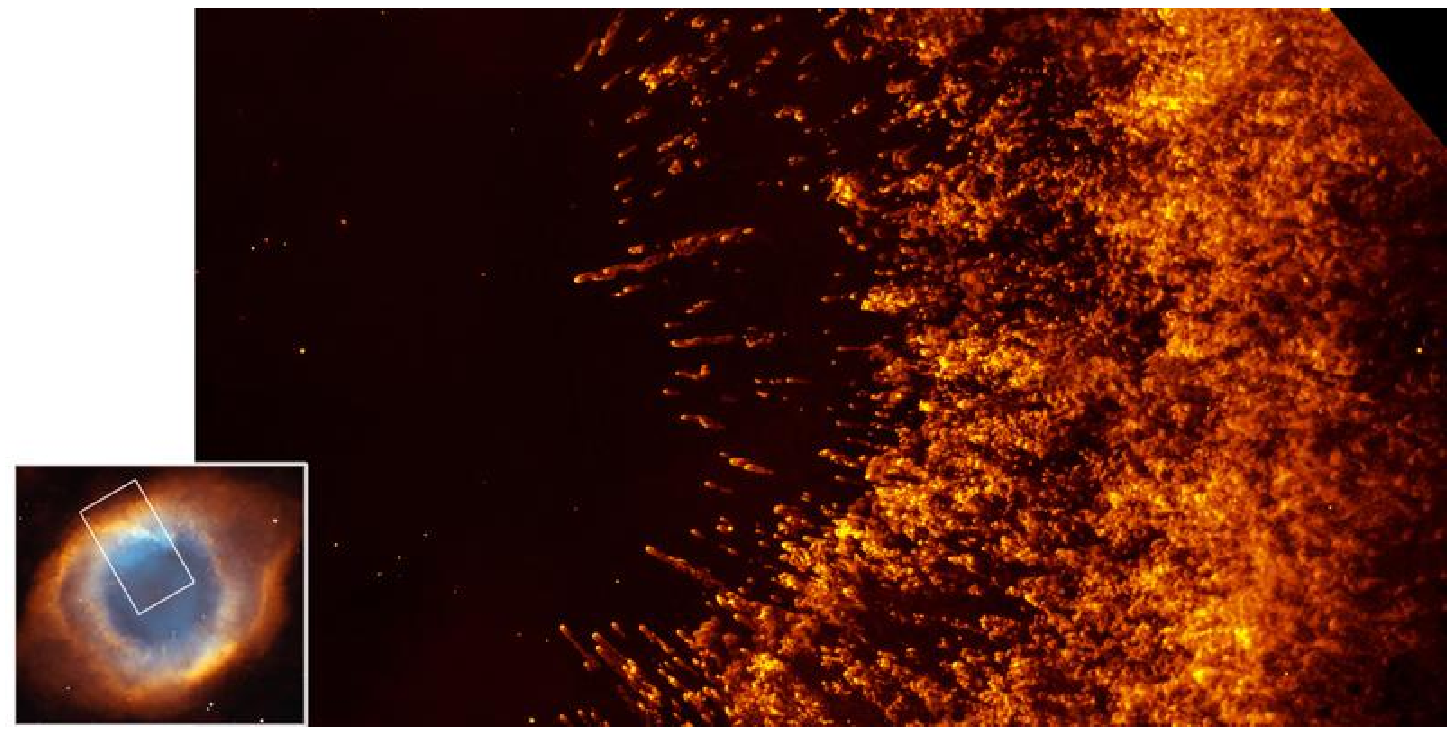,width=16cm,height=10cm} 
\caption{
An image of the Helix nebula obtained with the Subaru Telescope. 
At an approximate  distance of  700 light years, the Helix Nebula
is the closest example of a planetary nebula created at the 
end of the life of a Sun-like star.
The most striking feature of the Helix, first revealed by ground-based 
images, is its collection of more than 40,000 distinct gas blobs 
that resemble 
comets due to their compact heads and long, streaming tails. Each 
'cometary blob' is about twice
the size of our solar system and has about an Earth's-mass of 
hydrogen and other gases that were expelled from the nebula's central star 
thousands of years ago. Image Credit:
Matsuura, M.  et al. (National Astronomical Observatory of Japan).}
\label{Helix11}
\end{figure}

\begin{figure}[]
\centering
\epsfig{file=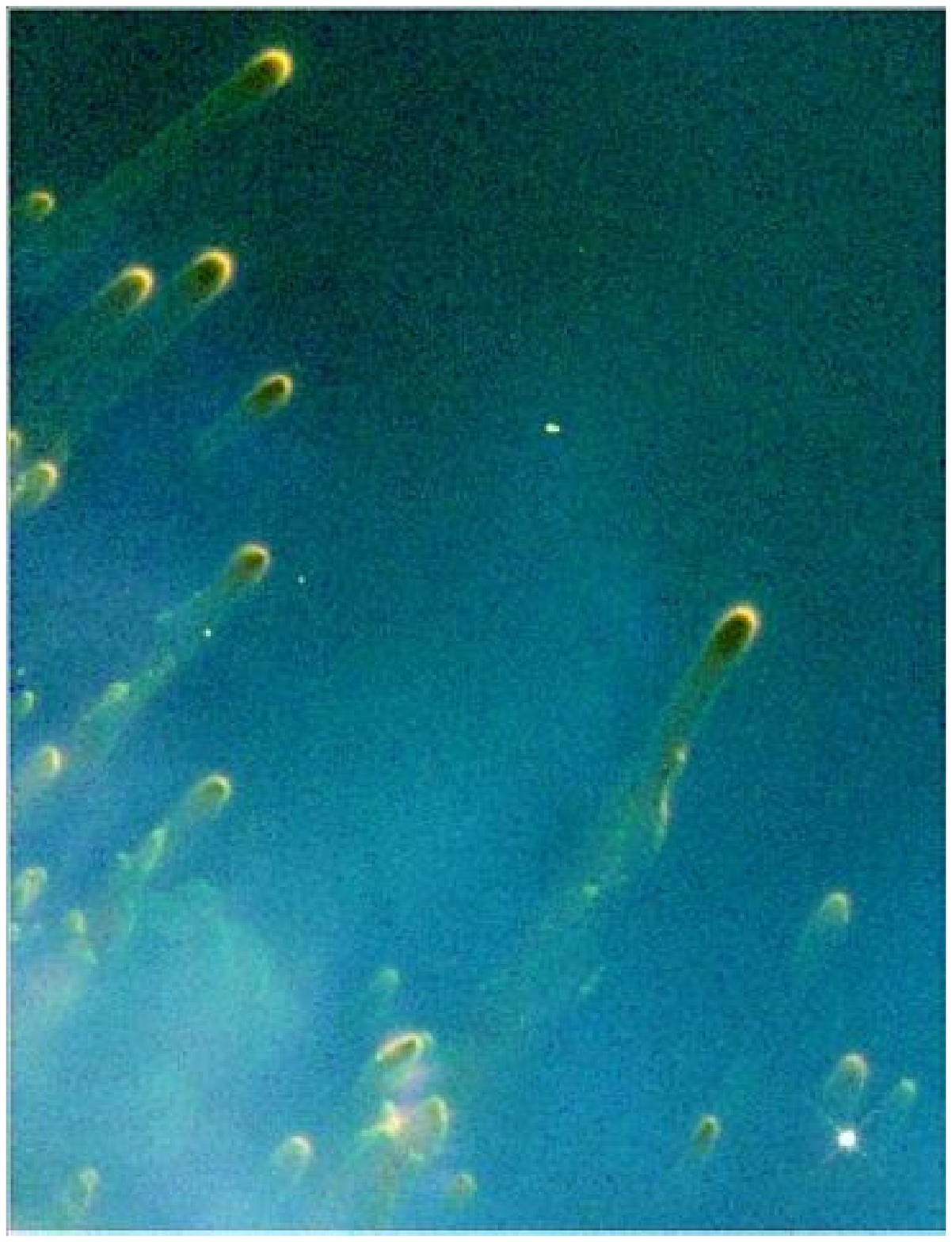,width=16cm,height=16cm} 
\caption{Hubble Space Telescope zoom on a section of the Helix Nebula
showing in great detail some of its cometary blobs. The tails 
of these 
gas blobs, which resemble the much smaller solar system comets, probably, 
were formed  by  fast winds from the central star
which overtook these blobs. Image Credit: Robert O'Dell, Kerry P. Handron 
(Rice University, Houston, Texas) and NASA.} 
\label{f2}
\end{figure}

\end{document}